\documentclass{article}
\usepackage[english]{babel}
\usepackage{graphicx}
\usepackage{amssymb,amsmath,fullpage,subfigure}
\usepackage{amsfonts,floatrow,wrapfig,multirow, adjustbox}
\usepackage{amsfonts,amsmath,amsthm,array,amssymb}
\usepackage{hyperref}
\usepackage{color}
\usepackage{xcolor}
\usepackage{mathrsfs}
\usepackage[english]{babel}

\include{latexsymb}
\include{pxfonts}
\include{displaymath,enumerate}
\include{tipa}

\begin{document}

 \begin{center}
 {\Large \textbf{Statistical Inference of Kumaraswamy distribution under  imprecise information}}
 \end{center}
 \bigskip
 \begin{center}
 {\bf \large Indranil Ghosh}\\ {\large University of North Carolina, Wilmington, North Carolina}\\
 e-mail: {\it  ghoshi@uncw.edu}\\
 \end{center}

 \begin{abstract}
 Traditional statistical approaches for estimating the parameters of the Kumaraswamy distribution  have dealt  with precise information. However, in real world situations, some information about an underlying experimental process might be imprecise and might be represented in the form of fuzzy information. In this paper, we consider the problem of estimating the parameters of a univariate Kumaraswamy  distribution  with two parameters when the available observations are described by means of fuzzy information. We derive  the maximum likelihood estimate of the parameters by using Newton- Raphson as well as EM algorithm method.  Furthermore, we  provide an approximation namely, Tierney and Kadane's approximation, to compute the Bayes estimates of the unknown parameters. The estimation procedures are discussed in details and compared via  Markov Chain Monte Carlo simulations in terms of their average biases and mean squared errors.
 \end{abstract}

 \section{Introduction}
  Kumaraswamy (1980) introduced a  two parameter absolutely continuous distribution which compares extremely favorably, in terms of simplicity, with the beta distribution.  The Kumaraswamy distribution on the interval $(0, 1)$, has
its probability density function (pdf) and its cumulative distribution function (cdf) with two shape parameters $a > 0$ and $b > 0$  defined by
\begin{equation}\label{eq1.1}
f(x)=a\ b x^{a-1}(1 - x^{a})^{b-1}I(0<x<1) \quad \text{and } F(x)=1-(1- x^{a})^{b}.
\end{equation}
If a random variable $X$ has pdf given in  (1)  then we will write $X\sim K(a,b).$

 The density function in (1) has similar properties to those of the beta distribution but has some advantages in terms of tractability.  The Kumaraswamy pdf is unimodal, uniantimodal, increasing, decreasing or constant depending (similar to the beta distribution) on the values of the parameters. It has some basic properties of the beta distribution: $a>1$ and $b >1$ (unimodal); $a <1$ and $b <1$ (uniantimodal); $a>1$ and $b \leq 1$ (increasing);  $a \leq 1$ and $b >1$ (decreasing); $a =b =1$ (constant). For a detailed survey of properties of the Kumaraswamy distribution, the reader is referred to Jones (2009). This distribution has a close relation with beta and generalized beta (first kind)  listed below:
 \begin{itemize}
 \item  If  $X \sim \textrm{Beta}(1,b)$ \,   then  $X \sim \textrm{K}(1,b)$

 \item  If   $X \sim \textrm{Beta}(a,1)$ \,   then  $X \sim \textrm{K}(a,1)$

 \item  If  $X \sim \textrm{K}(a,b),$  then  $X \sim \textrm{GB1}(a, 1, 1, b),$
 \end{itemize}

\noindent where GB1 stands for the generalized beta distribution of the first kind.

Over the last few years, there has been a great interest in studying the Kumaraswamy distribution, and mixing with other well-known probability models to achieve greater flexibility in modeling several types of real data exhibiting various patterns. For example, Nadarajah et al. (2012) studied a new generalized distribution by mixing Kumaraswamy distribution with an arbitrary  baseline $G$ distribution. Alizadeh et al. (2015) studied a new model by mixing Kumaraswamy with Marshall-Olkin family of distributions. In a separate article, Nadarajah et al. (2013) studied a mixture of Kumaraswamy and generalized Pareto model. Ghosh (2015) independently studied a Kumaraswamy mixture with Pareto (type IV) model useful for income modeling. Again, in another article, Ghosh (2014), derived and discussed another Kumaraswamy generalization, with mixing with a half-Cauchy distribution. Regarding discrete mixture, Ramos et al. (2015) developed  and  studied a new distribution, namely the Kumaraswamy-G Poisson family of distributions and discussed the associated inferences for the model parameters. Ghosh and Nadarajah (2016) studied in details, Bayesian inference for Kumaraswamy distribution based on censored samples.
All the above references are indicative of the fact that the Kumaraswamy  distribution has a greater applicability when it comes to modeling an observed phenomena, with possibly, the values of the variable of interest are somehow bounded between $(0,1)$.

However, majority of the inferential work for the Kumaraswamy distribution has been conducted under the assumption that complete data are available. In contrast, not much work has been done in the direction of missing (and or imprecise) information scenario with regard to inferential strategy for the Kumaraswamy distribution. This is a major motivation for this article.

It has been observed that  in numerous real life  situations we encounter  data which are not only random in nature  but ambiguous as well. It is to be noted that randomness involves only uncertainties in the outcomes of an experiment, while ambiguity, on the other hand, involves uncertainties in the meaning of the data. For  example, consider a case study on the  electric bulb manufacturing process that focuses on the lifetime of an electric bulb. An electric bulb may work perfectly over a certain period but may not work efficiently for some time, and finally becomes totally exhausted after  a certain time point. Therefore, the lifetime of each electric bulb may be reported by means of ambiguous statements such as ``approximately lower than 95 hours ", ``approximately 25 to 40 hours", ``approximately 74 to 98 but near to 110 hours ", ``approximately higher than 125 hours" and so on. In such a scenario, randomness occurs when the electric bulbs are selected at random and vagueness (or ambiguity)  is due to limited ability of the observer to describe the lifetime of those randomly selected electric bulbs using numbers. To deal with both types of uncertainties - randomness and vagueness, it is important  to incorporate fuzzy concept into statistical toolbox.

In recent years,  numerous  papers on generalization of classical statistical methods to analysis of fuzzy data have appeared in the literature. Wu (2004) discussed the Bayesian estimation on lifetime data under fuzzy environments. Gil et al. (2006) presented a backward analysis on the interpretation, modeling and impact of the concept of fuzzy random variable. Viertl (2006) studied generalization of classical statistical inference methods for univariate fuzzy data. Zarei et al. (2012) considered the Bayesian estimation of failure rate and mean time to failure based on vague set theory in the case of complete and censored data sets. Very recently, Pak et al. (2013, 2014) conducted a series of studies to develop the inferential procedures for the lifetime distributions on the basis of fuzzy data.

 The main objective  of this paper is to obtain the  suitable  inferential procedures for a Kumaraswamy distribution when the available observations are reported by means of fuzzy information.  We first describe the construction of fuzzy data from imprecise (equivalently vague) observations, and then discuss the computation of maximum likelihood estimate of the parameter $a$ and $b$.  Based on fuzzy data, there is no closed form for the MLE; therefore, we employ the EM algorithm to determine the maximum likelihood estimate. We also construct the approximate confidence interval of the unknown parameters by using the asymptotic distribution of the MLEs. Additionally, we consider the Bayesian inference of the parameters of the Kumaraswamy distribution. Since the Bayes estimates cannot be obtained in explicit form, we provide an approximation, namely Tierney and Kadanes approximation, as well as a Markov Chain Monte Carlo (MCMC) technique to compute those estimated and construct the highest posterior density (HPD) credible interval of the parameters $a $ and $b$.

 The rest of this paper is organized as follows. In Section 2, we obtain the maximum likelihood estimates of the parameters $a$ and $b$ , and also construct the approximate confidence intervals by using asymptotic normality of the MLEs. The Bayesian analyses are provided in Section 3. A Monte Carlo simulation study is presented in Section 4, which provides a comparison of all estimation procedures developed in this paper. Some concluding remarks are presented in Section 5.

In the following, at first, we consider the fundamental notation and some basic basic definitions of fuzzy set theory which will be frequently used in this paper. Consider an experiment characterized by a probability space $X=\left(\Omega, \mathscr{F}, \mathscr{P}_{\theta}\right),$  where  $ \left(\Omega, \mathscr{F}\right)$    is a Borel measurable space and $\mathscr{P}_\theta$  belongs to a specified family of probability measures $\left({\mathscr{P}_{\theta} ,\theta\in \Theta }\right)$ on  $ \left(\Omega, \mathscr{F}\right)$.  Assume that the observer cannot distinguish or transmit with exactness the outcome in the performance of  , but that rather the available observation may be described in terms of fuzzy information which is defined as follows. For details on this topic, see Tanaka et al. (1979).

 \begin{itemize}

 \item {\bf Definition 1}: A fuzzy event $x$ on  $X,$  characterized by a Borel measurable membership
 function  $\mu_{\tilde{x}}(x)$ from $X$ to $[0,1],$ where $\mu_{\tilde{x}}(x)$  represents the ``grade of membership" of $x$ to $\tilde{x},$  is
called fuzzy information associated with the experiment $X.$  The set consisting of all observable events from the experiment  $X$ determines a fuzzy information system    associated with it, which is defined as follows.

 \item {\bf Definition 2}:  A fuzzy information system (henceforth, in short f.i.s.) $\tilde{X}$ associated with
the experiment  ${\bf X}$ is a fuzzy partition with fuzzy events on $X,$ that is a finite set of fuzzy events on $X$ satisfying the orthogonality condition
 $$\sum_{\tilde{x}\in \tilde{X}}(x)=1,$$  for all $x\in X$. Alternatively,
 according to  Zadeh(1968), given the experiment  $X=\left(\Omega, \mathscr{F}, \mathscr{P}_{\theta}\right),$ $\theta\in \Theta $    and a  f.i.s. $\tilde{X}$
 associated with it, each probability measure $\mathscr{P}_{\theta}$ on $\left(\Omega, \mathscr{F}\right),$  induces a probability measure
on $\tilde{X}$ defined as follows:

  \item {\bf Definition 3}: The probability distribution on $\tilde{X}$ induced by $\mathscr{P}_{\theta}$  is the mapping $\mathscr{P}$  from $X$ to
$[0,1]$ such that

 \begin{equation}
 \mathscr{P}(\tilde{x})=\int_{X}\mu_{\tilde{x}}(x)d\mathscr{P}_{\theta}(x),
 \end{equation}

 for $\tilde{x}\in \tilde{X}$.
 In particular, the conditional density of a continuous random variable $U$ with p.d.f. $g(u)$ given the
fuzzy event $\tilde{A}$ can be defined as
 \begin{equation}
 g\left(u|\tilde{A}\right)
 =\frac{\mu_{\tilde{A}}(u)g(u)}{\int \mu_{\tilde{A}}(u)g(u)du}.
 \end{equation}

 For more details about the membership functions and probability measures of fuzzy sets, one can refer to Pak et al. (2013) and the references therein.
 In this context, we consider another definition due to Shafiq and Viertl (2014) which is as follows:

 \item {\bf Definition 4}: A fuzzy number is a subset, denoted by $\tilde{x}$, of
the set of real numbers (denoted by $\mathbb{R}$) and is characterized by the so called membership  function $\mu_{\tilde{x}}$, satisfying the following constraints:

\begin{enumerate}
\item [(i)] $\mu_{\tilde{x}}: \mathbb{R}\Longrightarrow [0,1]$ is Borel measurable;

\item [(ii)]  For every $x_{0}\in \mathbb{R}$, $\mu_{\tilde{x}}(x_{0})=1;$

 \item [(iii)] The usual $\lambda$-cuts ($0<\lambda\leq 1$), defined as $B_{\lambda}(\tilde{x})=\left\{x\in  \mathbb{R}: \mu_{\tilde{x}}(x)\geq \lambda\right\}, $are all closed interval, i.e., $_{\lambda}(\tilde{x})=\left[a_{\lambda}, b_{\lambda}\right], \quad \forall \lambda\in (0,1]$.

     \end{enumerate}

     Some widely known examples of membership functions to characterize fuzzy numbers are triangular and trapezoidal
fuzzy numbers. For example, triangular fuzzy number is defined as $\tilde{x}=\left(\xi, \omega, \delta\right)$ with the
corresponding membership function

\begin{equation*}
\mu_{\tilde{x}}
=\begin{cases}
\frac{x-\xi}{\omega-\xi}, \quad \xi\leq x\leq \omega,\\

\frac{\delta-x}{\delta-\omega}, \quad \omega\leq x\leq \delta,\\

0, \quad \text{elsewhere}.
\end{cases}
\end{equation*}

 Similarly, a   trapezoidal fuzzy number can be defined as ˜$\tilde{x} = \left(\xi, \omega, \delta, \theta\right)$ with the corresponding membership function

\begin{equation*}
\mu_{\tilde{x}}
=\begin{cases}
\frac{x-\xi}{\omega-\xi}, \quad \xi\leq x\leq \omega,\\

1, \quad \omega\leq x\leq \delta,\\

\frac{\theta-x}{\theta-\delta}, \quad \delta\leq x\leq \theta,\\

0, \quad \text{elsewhere}.
\end{cases}
\end{equation*}

 \end{itemize}
 
 \smallskip

\noindent Let us again revisit the example as mentioned earlier in the context of life length of an electric bulb.

{\bf Example 1:} Consider a life-testing experiment in which $n$ identical electric bulbs
(made by the same company) are placed on test. A tested electric bulb may be considered as failed, or to be more precise
nonconforming, when at least one value of its parameters (constituent parts) falls beyond specification
limits. In reality, however, the observer does not have the possibility to measure all
parameters and may not be able to define precisely the moment of a failure. So, he/she provides
an interval $[\xi_{i}, \omega_{i}]$  which certainly contains the lifetime of the electric bulb marked $i$ and an interval $[\xi_{i}, \theta_{i}]$ which
contains highly plausible values for that lifetime. This information may be encoded as a trapezoidal fuzzy number $\tilde{x}_{i} = \left(\xi_{i}, \omega_{i}, \delta_{i}, \theta_{i}\right)$ with the corresponding associated membership function

\begin{equation*}
\mu_{\tilde{x}_{i}}
=\begin{cases}
\frac{x_{i}-\xi_{i}}{\omega_{i}-\xi_{i}}, \quad \xi_{i}\leq x_{i}\leq \omega_{i},\\

1, \quad \omega_{i}\leq x_{i}\leq \delta_{i}\\

\frac{\theta_{i}-x_{i}}{\theta_{i}-\delta_{i}}, \quad \delta_{i}\leq x_{i}\leq \theta_{i},\\

0, \quad \text{elsewhere}.
\end{cases}
\end{equation*}

In this case randomness arises from the selection of electric bulbs as well as other observable factors
which influence the perception by the observer. In contrast, fuzziness arises from the
meaning of the reported failure times.

 \section{Maximum likelihood estimation}
 Suppose that   $X_{1} ,\cdots,  X_{n}$ is a random sample of size $n$ from Kumaraswamy distribution with the density function
given by (1).   Let     $\underline{X}=\left( X_{1} ,\cdots,  X_{n} \right)$ denotes the corresponding random vector. If a realization $\underline{x}$ of $\underline{X}$
was known exactly, one can obtain the complete-data log-likelihood function as

 \begin{equation}
 \log L\left(\underline{x}; a,b\right)
 =n\left(\log a+\log b\right)+(a-1)\sum_{i=1}^{n}\log x_{i} +(b-1)\sum_{i=1}^{n}\log\left(1-x^{a}_{i}\right).
 \end{equation}

 \noindent Next,  consider the situation  where  $\underline{x}$  is not observed precisely, and only partial information about $\underline{x}$
is available in the form of fuzzy observation $\tilde{\underline{x}}=\left( \tilde{\underline{x}_{1}}, \tilde{\underline{x}_{2}}, \cdots, \tilde{\underline{x}_{n}}\right)$ with the Borel measurable membership
function  $\mu_{\tilde{x}}$.  In reality, the grade of membership $\mu_{\tilde{x}}(\underline{x})$  is often regarded as a
``probability with which the observer gets the information $\tilde{\underline{x}}$ when he/she really has obtained the exact
outcome $\underline{x}$".
Once $\displaystyle\tilde{\underline{x}}$ is given, we can obtain the observed data log-likelihood function by using the expression
(4) as follows:

  \begin{equation}
\ell_{0}= \log L\left(\tilde{\underline{x}}; a,b\right)
=\displaystyle n\left(\log a+\log b\right)+(a-1)\sum_{i=1}^{n}\log\int x_{i} \mu_{\tilde{x_{i}}}(x) dx +(b-1)\sum_{i=1}^{n}\log \int \left(1-x^{a}_{i}\right)\mu_{\tilde{x_{i}}}(x) dx.
 \end{equation}
The maximum likelihood estimate of the parameters $a$  and  $b$  can be obtained by maximizing the
log-likelihood $\ell_{0}$.  Equating the partial derivatives of the log-likelihood (5) with respect to
 $a$ and  $b$  to zero, the resulting maximum likelihood equations are:

\begin{equation}
\frac{\partial \ell_{0}}{\partial a}
=\displaystyle\frac{n}{a}+\sum_{i=1}^{n}\log\int x_{i} \mu_{\tilde{x_{i}}}(x) dx-(b-1)\sum_{i=1}^{n}\frac{\int x^{a}_{i}\log(x_{i})\mu_{\tilde{x_{i}}}(x) dx }{\int \left(1-x^{a}_{i}\right)\mu_{\tilde{x_{i}}}(x) dx}.
\end{equation}

\begin{equation}
\frac{\partial \ell_{0}}{\partial b}
=\displaystyle\frac{n}{b}+\sum_{i=1}^{n}\log \int \left(1-x^{a}_{i}\right)\mu_{\tilde{x_{i}}}(x) dx.
\end{equation}

 Since there are no closed form of the solutions to the likelihood equations (6) and (7), an iterative numerical search  procedure needs to be considered  to obtain the MLEs. Next, we describe two widely practiced  search procedures, namely,   the Newton- Raphson method and the EM algorithm to determine the MLEs of the parameters $a$  and  $b$.

\subsection{Newton- Raphson procedure}
Newton-Raphson algorithm is a direct approach for estimating the relevant parameters in a likelihood function. In this procedure, the solution of the likelihood equation is obtained through an iterative procedure  which is as follows. Let $\underline{\delta}=\left(a,b\right)^{T}$, be the parameter vector, where $T$ stands for transpose.
Next, at the  $(h +1)$ th step of iteration process, the updated parameter is obtained as

\begin{equation}
\delta^{h+1}
=\delta^{h} -\left[\frac{\partial \ell_{0}}{\partial \delta}|_{\delta=\delta^{h}}\right]^{T}
\times \left[\frac{\partial \ell_{0}}{\partial \delta\partial \delta^{T}}|_{\delta=\delta^{h}}\right]^{-1},
\end{equation}

\noindent where

\begin{equation*}
\frac{\partial \ell_{0}}{\partial \underline{\delta}}
=\begin{pmatrix}
\frac{\partial \ell_{0}}{\partial a} \\

\\

\frac{\partial \ell_{0}}{\partial b}
\end{pmatrix}.
\end{equation*}

 And,

 \begin{equation*}
\frac{\partial^{2} \ell_{0}}{\partial \underline{\delta}  \underline{\delta}^{T}}
=\begin{pmatrix}

 \frac{\partial^{2} \ell_{0}}{\partial a^{2}} & \frac{\partial^{2} \ell_{0}}{\partial a \partial b}\\

 \\
 \frac{\partial^{2} \ell_{0}}{\partial a \partial b} & \frac{\partial^{2} \ell_{0}}{\partial b^{2}}
 \end{pmatrix},
\end{equation*}

 \noindent where  the second-order derivatives of the log-likelihood with respect to the parameters, required for proceeding with the Newton- Raphson method, are obtained as follows:

\begin{equation*}
 \frac{\partial^{2} \ell_{0}}{\partial a^{2}}
 =-\frac{n}{a^{2}}-(b-1)\sum_{i=1}^{n} \frac{\int  x^{a}_{i}\left(\log(x_{i})\right)^{2}\mu_{\tilde{x_{i}}}(x) dx }{\int \left(1-x^{a}_{i}\right)\mu_{\tilde{x_{i}}}(x) dx}.
 \end{equation*}

 \begin{equation*}
 \frac{\partial^{2} \ell_{0}}{\partial b^{2}}
 =-\frac{n}{b^{2}}.
 \end{equation*}

 \begin{equation*}
 \frac{\partial^{2} \ell_{0}}{\partial a \partial b}
 =-\sum_{i=1}^{n} \frac{\int  x^{a}_{i}\left(\log(x_{i})\right)\mu_{\tilde{x_{i}}}(x) dx }{\int \left(1-x^{a}_{i}\right)\mu_{\tilde{x_{i}}}(x) dx}.
 \end{equation*}

 The iteration process then continues until convergence, i.e., until    $|| \underline{\delta}^{h+1} -  \underline{\delta}^{h}||\leq \epsilon,$  for some predefined $\epsilon >0$. Note  that the second-order derivatives of the log-likelihood are required at every iteration stage in the Newton- Raphson method. However, quite often, the computation of the derivatives based on fuzzy data can be  really troublesome. This is major drawback of this method.

To remedy against this melody, a  viable alternative to the Newton- Raphson algorithm is the well-known EM algorithm. In the following, we discuss how that can be utilized  to determine the MLEs in this case.

\subsection{EM algorithm}

The Expectation Maximization (EM) algorithm is a widely applicable approach to the iterative
 computation of maximum likelihood estimates and useful in a variety of incomplete-data problems. For details, see Dempster  et al. (1977).
Since the observed fuzzy data x can be seen as an incomplete specification of a complete data vector $x$, the EM algorithm is applicable to obtain the maximum likelihood estimates of the unknown parameters. In the following, we use the EM algorithm to determine the MLEs of  $a$ and
$b$.

 Based on complete data log-likelihood function from (4),  and taking the  partial derivative with respect to $a$  and $b$, respectively, the following likelihood  equations are obtained as follows:

\begin{equation}
\frac{n}{a}
=- \displaystyle \log x_{i}.
\end{equation}

\begin{equation}
\frac{n}{b}
=-\displaystyle \sum_{i=1}^{n}\log\left(1-x^{a}_{i}\right).
\end{equation}

 \noindent Therefore the EM algorithm is given by the following iterative process

 \begin{enumerate}
 \item  Start with an initial starting given values of $a$ and $b$, say, $a^{(0)}$ and  $b^{(0)}$ and set $h=0$.

 \item At the  $(h +1)$ th stage of  iteration:

 \begin{itemize}
 \item The E-step requires to compute the following conditional expectations using the expression (5):

 \begin{equation*}
 E_{a^{(h)}, b^{(h)}}\left[\log \underline{X}| \tilde{x}\right]
 =-\displaystyle \frac{\int \log \left(x\right) x^{-(a^{h}+1)}\mu_{\tilde{x_{i}}}(x) dx}{\int x^{-(a^{h}+1)}\mu_{\tilde{x_{i}}}(x) dx}.
 \end{equation*}

  \begin{equation*}
 E_{a^{(h)}, b^{(h)}}\left[\log\left(1-X^{a^{h}}\right) | \tilde{x}\right]
 =-\displaystyle \frac{\int \log \left(1-x^{a^{h}}\right)^{-(a^{h}+1)}\mu_{\tilde{x_{i}}}(x) dx}{\int\left(1-x^{a^{h}}\right)^{-(a^{h}+1)}\mu_{\tilde{x_{i}}}(x) dx}. \end{equation*}

\end{itemize}

 \end{enumerate}

\section{Bayesian estimation}
In this section we describe the Bayes estimate of the unknown parameter as well as the
corresponding highest posterior density credible interval. In the Bayesian estimation unknown
parameter is assumed to behave as random variable with distribution commonly
known as prior probability distribution.
 Here,  we  consider the following independent gamma priors for all the  parameters $a,$  $b$ given as follows:
\begin{itemize}
\item Prior for $a$ : $\Pi(a)\sim \Gamma(2.1, 1.7)$.

\item Prior for $b$ : $\Pi(b)\sim \Gamma(1.3, 0.89)$.
\end{itemize}

\bigskip

{\bf Note:} We do not claim that these choices of the hyperparameters are the optimal or uniformly best in all situations like this. However, in all the simulations/examples tried, we found this to be a reasonable one. Of course there might be others.

\bigskip

By combining (5) with the above set of independent priors, the joint density function of the data
and the parameters $a$ and $b$ becomes

\begin{eqnarray}
&&\Pi(data, a,b)\propto \displaystyle \left(a^{1.4}b^{0.3}\exp\left(-\left[1.8 a+1.7 b\right]\right)\right)\notag\\
&&\times n\left(\log a+\log b\right)+(a-1)\sum_{i=1}^{n}\log\int x_{i} \mu_{\tilde{x_{i}}}(x) dx +(b-1)\sum_{i=1}^{n}\log \int \left(1-x^{a}_{i}\right)\mu_{\tilde{x_{i}}}(x) dx.
\end{eqnarray}

Therefore, the marginal posterior density functions of $a$ (and $b$) respectively  given the data can be obtained as

\begin{itemize}

\item $\Pi(a|data)\propto \int_{0}^{\infty}\Pi(data, a,b) db.$

\item $\Pi(b|data)\propto \int_{0}^{\infty}\Pi(data, a,b) da.$
\end{itemize}

\bigskip

Note that the Bayes estimate of any function of $a$, say $h(a),$ under squared error
loss function is the posterior mean which is given by
\begin{equation}
\int_{0}^{\infty}\Pi(a|data)h(a) da.
\end{equation}

and similarly for the other parameter $b$ as well.

However, the Equations (11) and (12) are not available in analytically tractable and closed nice form due to the complex form of
the likelihood function. Therefore, we use Tierney and Kadane’s approximation as well as
MCMC method for computing the Bayes estimate of $a$ and $b$.

\subsection{Tierney and Kadane's approximation}

First, we rewrite the expression in (11) as (for both the parameters $a$ and $b$ respectively)
\begin{equation}
\int_{0}^{\infty}\Pi(a|data)h(a) da
=\frac{\int_{0}^{\infty}\exp\left(n F^{*}(a)\right) da}{\int_{0}^{\infty}\exp\left(n F(a)\right)da},
\end{equation}

and

\begin{equation}
\int_{0}^{\infty}\Pi(b|data)h(b) db
=\frac{\int_{0}^{\infty}\exp\left(n F^{*}(b)\right) db}{\int_{0}^{\infty}\exp\left(n F(b)\right)db},
\end{equation}

\bigskip

\noindent where $$F(a)=\frac{1}{n}\log \Pi(data, a),$$ and $$F^{*}(a)=F(a)+\frac{1}{n}\log h(a).$$ Tierney and Kadane (1986) applied Laplace’s method to produce an approximation of
(17) as follows:
\begin{equation}
\hat{h}_{BT}(a)=\left[\frac{\theta^{*}}{\theta}\right]^{1/2}\exp\left(n\left[F^{*}(\bar{a^{*}})-F(\bar{a})\right]\right),
\end{equation}

\noindent where $\bar{a^{*}}$ and $\bar{a}$ maximize $F^{*}(\bar{a^{*}})$ and $F(\bar{a})$, respectively, and $\theta^{*}$ and $\theta$ are minus of the inverse of the second derivatives of $F^{*}(a)$ and  $F(a)$ at $\bar{a^{*}}$ and $\bar{a}$ respectively.

Similar operation will be assumed for the other parameter $b$ as well.   Next, we apply this approximation to obtain the Bayes estimate of the parameter $a$. Setting $h(a) = a,$ we have

\begin{eqnarray}
F(a)
&=& \frac{1}{n}\int_{0}^{\infty} \left\{ 1.4 \log a-(1.8 a+1.7 b)+ 0.3 \log b+ n\left(\log a+\log b\right)+(a-1)\sum_{i=1}^{n}\log\int x_{i} \mu_{\tilde{x_{i}}}(x) dx\right.\notag\\
&&\left. +(b-1)\sum_{i=1}^{n}\log \int \left(1-x^{a}_{i}\right)\mu_{\tilde{x_{i}}}(x) dx\right\}db,
\end{eqnarray}

and

\begin{eqnarray}
F^{*}(a)
&=& \frac{1}{n}\int_{0}^{\infty} \left\{ 2.4 \log a-(1.8 a+1.7 b)+ 0.3 \log b+ n\left(\log a+\log b\right)+(a-1)\sum_{i=1}^{n}\log\int x_{i} \mu_{\tilde{x_{i}}}(x) dx\right.\notag\\
&&\left. +(b-1)\sum_{i=1}^{n}\log \int \left(1-x^{a}_{i}\right)\mu_{\tilde{x_{i}}}(x) dx\right\}db.
\end{eqnarray}

On substitution of (16) and (17) in (15), one can obtain the Bayes estimate of $a$ under squared error loss.  Similar approach can also be made to obtain the Bayes estimate of $b$ under square error loss.

\subsection{MCMC and HPD credible interval}
Here, we first draw random samples from the posterior density function (15).
Then, we compute the Bayes estimates of $a$ and $b$  and also construct its Highest Posterior Density (in short HPD) credible interval.
Since the joint posterior density function in (11)  can not be computed explicitly, we use a Metropolis-
Hastings algorithm to generate samples from posterior density of $\tau=\left(a,b\right)$ as follows:

The Metropolis-Hastings algorithm is carried out considering the following steps:
\begin{enumerate}
\item We consider a starting (initial) value $\underline{\tau}^{(0)}=\left(a^{(0)}, b^{(0)}\right)$.

\item At iteration stage $t$, draw $\underline\tau^{(*)}$ from a jumping distribution $J_{t}(\underline{\tau}|\underline{\tau}^{(t-1)})$.

\item Compute an acceptance ratio
$r=\displaystyle\frac{f(\underline\tau^{(*)}|data)/J_{t}(\underline{\tau}|\underline{\tau}^{(t-1)})}
{f(\underline{\tau}^{(t-1)}|data)/J_{t}(\underline{\tau}^{(t-1)}|\underline{\tau}^{(*)})}$.

\item Accept  $\underline\tau^{(*)}$ with probability min $(r,1)$. If $\underline\tau^{(*)}$ is not accepted then
$\underline{\tau}^{(t)}=\underline{\tau}^{(t-1)}$.

\item We repeat steps $2-4$ , $M$ times to get $M$ draws from $f(\underline{\tau}|data)$. We consider $20$ parallel chains of length $M$ each.  Thus, we will have $a_{j}$ and $b_{j}$ for $j=1,2,\cdots,M$ in general.

\end{enumerate}

\smallskip

The retained sample values, say, $\tau_{1},\cdots, \tau_{M}$, are a random sample from the joint posterior density
$π(a,b | data).$ Next, by using Monte Carlo integration technique Rubinstein and Kroese
(2006), the Bayes estimate of $a$ and $b$ under squared error loss function can be obtained as

$$a_{BM}=\frac{1}{M}\sum_{i=1}^{M}a_{i},$$ and $$b_{BM}=\frac{1}{M}\sum_{i=1}^{M}b_{i}.$$ For constructing HPD credible interval of $a$ (say), one may use the method proposed by Chen and
Shao (1999) as follows:

\noindent Let $a_{(1)}<\cdots<a_{(M)}$ be the ordered values of $a_{i}$ for $i=1,2,\cdots, M$. Then, consider the following $100(1-\alpha)\%$ credible intervals of $a$:

$$\left(a_{(1)}, a_{((1-\alpha)M)}\right), \cdots, \left(a_{(\alpha M)}, a_{(M)}\right).$$

The HPD credible interval of $a$ can be derived by choosing the interval which has the
shortest length.

Similarly, one may obtain the HPD interval for the parameter $b$.

 \section{Simulation study}
In this section, simulation studies are conducted to compare the performances of the different
estimators and also different confidence/credible intervals. Our main objective is to compare the
performances of the MLE and Bayes estimates of the unknown parameters, in terms of
their average values and mean squared errors. We also compare the average lengths of the
asymptotic confidence intervals to the HPD credible intervals and their coverage percentages.
All the computations are performed on R-programming environment. For simulation purposes, we have considered $a=2,$ and $b=3$ and different choices of sample sizes,
namely $ n = 50, 75, 100, 150, 200.$ For each $n,$ we have generated random sample from the Kumaraswamy distribution with $a=2, b=3$. Then, using the method as proposed by Pak et al. (2014), each realization
of the generated samples was fuzzified by employing fuzzy information system.  The estimates of the parameters $a$ and $b$ for the fuzzy sample were computed using the maximum
likelihood method and under the Bayesian paradigm (with independent priors set up). For initial choices of the parameters ($a, b$) required for the MLE method, we have taken values that are wide apart from the actual values of the parameters. For computing the Bayes estimate, we
have assumed that both $a$ and $b$ have independent gamma priors with specific choices of the hyperparameters (described earlier). We replicate the process 20000
times with a burn in of $1000$ samples  and report the average values (AV) and mean squared errors (MSE) of the estimates
in Tables 1-2.

Furthermore, we provide an approximate 95\% confidence interval and also the HPD credible
interval of the unknown parameters. Criteria appropriate to the evaluation of the
two methods under consideration include: closeness of the coverage probability to its nominal
value and expected interval width. For each simulated sample, we have computed confidence/
credible intervals and checked whether the true value of the parameter lay within the  intervals and recorded the length of the intervals. The estimated coverage probability was
computed as the number of intervals that covered the true value divided by $20000$ while
the estimated expected width of the intervals was computed as the sum of the lengths
for all intervals divided by $20000.$ The coverage probabilities and the expected widths for
different sample sizes are presented in Tables 1-2.

From Tables 1-2, one may observe the following:
\begin{itemize}
 \item The MSE of the estimators decrease
significantly as the sample size n increases, as one would expected.

\item  The performances of
the Bayes estimates with  with informative prior are uniformly better. It is also seen that the Bayes
estimates obtained by Tierney and Kadane’s approximation and the MCMC method behave
in a similar manner. So, we can not say that one procedure is uniformly better than the other (while comparing Tierney and Kadane’s approximation and the MCMC method). It should be noted here that although the MCMC techniques are
computationally expensive, but in turn we can use them to construct HPD credible interval. 
\end{itemize}

Next, considering the confidence and credible intervals, it is observed that the asymptotic
results of the MLE work quite satisfactorily. It can maintain the coverage percentages in most
of the cases even when the sample size is relatively small. The widths of the confidence/credible
intervals decreases with an increase in the sample size $n$ as expected. The performances of the
credible intervals are quite good and their coverage percentages are close to the corresponding
nominal level. Moreover, in most of the cases, the average lengths of the credible
intervals are slightly shorter than the confidence intervals. 
One may consider a dependent prior choice set up for the Bayesian inference and perform a similar study.

 \section{Conclusions}
   A lot of work has been done regarding the estimation of parameters of the Kumaraswamy distributions based on complete and censored samples (see, for example, Ghosh and Nadarajah (2016)). However, in almost all cited references in this article, it was assumed that the available data are (or can be) obtained in exact numbers. In contrast, in many real world observed phenomena, the data obtained as an outcome of an experiment may not always be recorded/ evaluated/measured properly. As a consequence, there is a greater need of developing an appropriate statistical methodology to tackle such data and conduct a proper statistical analysis.

   In this paper, we have discussed several estimation procedures for the Kumaraswamy distribution when the reported data are available in the form of fuzzy information. In particular, we have discussed the traditional maximum likelihood method and the Bayesian procedure (both under the independent non-informative prior and dependent prior set up). From the simulation study, it appears that the performance of the MLE based on NR method is less efficient as compared to the EM algorithm. Although, one can not say in some absolute sense that one method is superior than the other always, but still the EM algorithm is preferred due to its computational simplicity. In terms of overall comparison (with respect to minimum average bias and MSEs) the performance of the Bayes estimates is generally best.

 \section*{Appendix}

 \begin{center}
Table 1: Averages values and mean squared errors of the ML estimates of $a$ and $b$, coverage
probabilities and expected width of 95\% confidence interval for different sample sizes.
\noindent\adjustbox{max width=\textwidth}{%
\begin{tabular}{p{1.1\textwidth}}
\[\begin{array}{|c|c|cc|cc|}
\hline
\textrm{Sample Size}&\textrm{Estimation method} &\multicolumn{2}{|c|}{\textrm{Bias \& (MSE)}}& \multicolumn{2}{|c|}{\textrm{CI for a \& CI for b}} \\
\hline
n & & \hat{a} & \hat{b}& \textrm{Coverage (width) a} &\textrm{Coverage (width)b} \\
\hline
50&\textrm{NR}&0.473(0.823)&0.379(0.768)&0.8231(0.4522) &0.8312(0.4655)\\	
    & \textrm{EM algorithm}&0.417( 0.624)&0.347(0.478)&0.8326(0.3765) &0.8411(0.4624)
      \\ \hline
75 &\textrm{NR}& 0.386(0.652)&0.361(0.533)&0.8843(0.3102) &0.8593(0.3318)\\	
    & \textrm{EM algorithm}& 0.316(0.533)&0.259(0.375)&0.8937(0.2958) &0.8642(0.3429)
     \\ \hline

100 &\textrm{NR}&0.326(0.437) &0.311(0.427)&0.9032(0.2648) &0.8947(0.2585)\\	
    &\textrm{EM algorithm}&0.289(0.417)&0.238(0.329)&0.9119(0.2215) &0.9067(0.2653)
     \\ \hline

 150 &\textrm{NR}&0.277(0.369) &0.254(0.322)&0.9245(0.2138) &0.9136(0.2484)\\	
    &\textrm{EM algorithm}&0.273(0.338)&0.214(0.243)&0.9307(0.2016) &0.9177(0.2152)
     \\ \hline
 200 &\textrm{NR}&0.226(0.304) &0.236(0.228)&0.9377(0.1868) &0.9435(0.1946)\\	
    &\textrm{EM algorithm}&0.213(0.195)&0.183(0.205)&0.9587(0.1773) &0.9513(0.1804)
     \\ \hline
\end{array}\]
\end{tabular}}
\end{center}

\newpage
\begin{center}
Table 2: Averages values and mean squared errors of the Bayes estimates of $a$ and $b$, coverage
probabilities and expected width of 95\% credible interval for different sample sizes.
\noindent\adjustbox{max width=\textwidth}{%
\begin{tabular}{p{1.1\textwidth}}
\[\begin{array}{|c|cc|cc|cc|}
\hline
\textrm{Sample Size} &\multicolumn{2}{|c|}{\textrm{Bias \& (MSE)under MCMC}}& \multicolumn{2}{|c|}{\textrm{Bias \& (MSE) under TK}}& \multicolumn{2}{|c|}{\textrm{CI for a \& CI for b}} \\
\hline
n &  \hat{a} & \hat{b}&   \hat{a} & \hat{b} & \textrm{Coverage (width) a} &\textrm{Coverage (width)b} \\
\hline
50&0.3323(0.368) &0.3145(0.326)&0.2861(0.657)&0.3218(0.387) &0.8455(0.3135) &0.8599(0.2950)\\	   
  \hline
75 &0.0643(0.312)& 0.1436(0.275)&0.2615(0.533)&0.1498(0.293) &0.8668(0.2917) &0.8718(0.2898)\\	
     \hline
100 &0.0327(0.265)&0.0224(0.281) &0.1749(0.427)&0.0263(0.286) &0.9177(0.2438) &0.9185(0.2393)\\	
    \hline

 150 &0.0205(0.218)&0.0115(0.219) &0.1232(0.322)&0.0127(0.244) &0.9253(0.2001) &0.9297(0.1963)\\	
     \hline
 200 &0.0162(0.175)&0.0069(0.183) &0.0976(0.228)&0.0112(0.198) &0.9614(0.1793) &0.9633(0.1762)\\	
     \hline
\end{array}\]
\end{tabular}}
\end{center}

 \end{document}